 \documentclass[10pt,final,twocolumn]{IEEEtran}

\usepackage{amsfonts}
\usepackage{amsmath}
\usepackage{setspace}
\usepackage{enumerate}
\usepackage{verbatim}
\usepackage{amssymb}
\usepackage{amsbsy}
\usepackage{theorem}
\usepackage{setspace}
\usepackage{url}

\usepackage[dvips]{epsfig}

\def\urltilde{\kern -.15em\lower .7ex\hbox{\~{}}\kern .04em}
\def\urldot{\kern -.10em.\kern -.10em}

\def\urlhttp{http\kern -.10em\lower -.1ex\hbox{:}\kern -.12em\lower 0ex\hbox{/}\kern -.18em\lower 0ex\hbox{/}}

\newtheorem{theorem}{Theorem}

\newtheorem{proposition}[theorem]{Proposition}

\newtheorem{Problem}{Problem}

\newcommand {\R}{\mathbb R}

\newcommand{\be}{\begin{equation}}
\newcommand{\ee}{\end{equation}}
\newcommand{\Int}{\operatorname{{\mathrm int}}}

{\theorembodyfont{\rmfamily} }

\newcommand{\LMD}{\lambda_0,\dots,\lambda_n}

\begin{document}
\title{Analyzing Linear Communication Networks using the Ribosome Flow Model\thanks{This research is partially supported by a  research grant
from  the Israeli
Ministry of Science, Technology \& Space.}
}
\author{
  Yoram Zarai, Oz Mendel, and Michael Margaliot
\thanks{Corresponding author: Prof. Michael Margaliot, School of Elec. Eng.-Systems, Tel Aviv University, Israel 69978. Homepage: {\tt www.eng.tau.ac.il/\urltilde michaelm} \;\; Email: {\tt michaelm@eng.tau.ac.il}  }  }
\maketitle

\begin{abstract}

The Ribosome Flow Model~(RFM) describes
 the unidirectional movement of interacting
  particles along a one-dimensional chain of sites.
 As a site becomes fuller, the effective entry rate into this site decreases.
 The~RFM has been used to model and analyze  mRNA  translation, a biological
 process in which  ribosomes (the particles)
 move along the mRNA molecule (the chain), and decode the genetic information into proteins.

Here we  propose the RFM as an analytical framework for
 modeling and analyzing linear communication networks.
In this context, the moving particles are data-packets,
   the chain of sites is a one dimensional set of ordered buffers,
   and the decreasing entry rate to a fuller buffer represents a kind of decentralized
	backpressure flow control.
   For an RFM with homogeneous 
   link capacities, we provide closed-form expressions for important network metrics including the
   throughput and end-to-end delay. We use these results to
    analyze the hop length and the transmission probability (in a contention access mode) that minimize the end-to-end delay in a multihop linear network,
     and provide closed-form expressions for the optimal parameter values.
       \end{abstract}


\section{Introduction}\label{sec:Intro}

The inflation of  wireless devices
and the dramatic increase in demand for real-time applications
 has led to
a growing interest in the analysis
 and design of  mobile ad hoc networks~(MANETs).  Important features of
  MANETs include: nodes may operate as clients  and as routers;  flow control is decentralized; and
  the communication infrastructure   between the nodes changes
   dynamically on the basis of network connectivity.
   The     end-to-end features of such networks
  depend  on the intricate  correlations between
  the different flows in the system.
  The  rigorous  analysis of  basic
  network characteristics, such as end-to-end delay, buffer utilization, and throughput,
  is a non-trivial  challenge

    Wireless   networks have been analyzed using different tools including
    information theory~\cite{network_info_book} and queuing theory~\cite{xie2009towards,Bisnik200979}.
     However,  these often use simplified assumptions and yield approximate results when characterizing the network  throughput and delay performances.
For example, large scale $2D$ networks have been studied based on  Kleinrock's Independence Assumption \cite{nelson1985spatial}. According to
this assumption, traffic correlations between different flows can be ignored and the effect on
delay performance is negligible, subject to sufficient traffic mixing and
moderate-to-heavy traffic loads (see e.g.,~\cite{delay_ana_ad_hoc}).
 However, recent studies suggest that this assumption does not hold in
 various real-world communication networks~\cite{xie2009towards}.
Ad-hoc networks have also been modeled as a network of Jackson queues~\cite{Kleinrock:1975:TVQ:1096491,jackson1957}. This is a stochastic model that assumes independent M/M/1 queues along the network. The service times are exponentially distributed and are mutually independent.  As such, the average number of data-packets in each node is determined independently from the node's neighbors and thus the interaction between adjacent nodes is   neglected.

 Andrews et al.~\cite{rethinking_info} point out that    standard analysis tools
 may not be adequate for    MANETs. They suggest several other tools, including non-equilibrium statistical
  mechanics. They state that:
``In particular, the microscopic theories of vehicular traffic are promising since they
do not treat vehicle flows as compressible fluids but explicitly focus on the dynamics of the
individual vehicle. In other words, the theories treat the vehicular network as a system of
interacting particles driven far from equilibrium, enabling the study of the dynamics of more
general non-equilibrium systems-such as MANETs.''

An interesting paper~\cite{DBLP:journals/tmc/SrinivasaH12} (see also~\cite{liutowards,conf/icccn/SrinivasaH10})
has  suggested using
 the \emph{Totally Asymmetric Simple Exclusion Process}~(TASEP)~\cite{TASEP_book}
 to model and analyze ad hoc networks.
   TASEP is a stochastic model for particles that move along some kind of ``tracks'' or ``trails''. A lattice of sites models the tracks.
    Each site may either be free or contain a single particle.
    Particles hop, with some probability, from one site to the consecutive  one,
    but only if the  target site is free (hence the term \emph{simple exclusion}). This models  
		particles that have volume and cannot overtake one another. In this way, the model
   captures the interaction between the particles.
   The term \emph{totally asymmetric} is used to indicate that  motion is unidirectional.

 TASEP and its variants
 are regarded
  as  a paradigmatic model in  non-equilibrium statistical mechanics,
 and have  been used to  model and  analyze numerous   natural and artificial
multiagent   systems, including
traffic flow, molecular motors, mRNA translation and more~\cite{nems2011,Chowdhury2005,TASEP_book}.
For example, in the context of traffic flow, the particles [lattice] represent  cars [the road].


Recently, Reuveni et al.~\cite{reuveni2011genome} studied a deterministic non-linear model, called the \emph{ribosome flow model}~(RFM), that can also be derived by mean-field approximation of TASEP.
Although nonlinear, the RFM is amenable to mathematical
analysis   using various tools including the
analytical theory of continued fractions~\cite{10.1109/TCBB.2012.120,infi_HRFM}, monotone systems
theory~\cite{margaliot2012stability,Margaliot06082013},
contraction theory~\cite{RFM_entrain},      convex optimization theory~\cite{RFM_concave},
and matrix theory~\cite{RFM_sense}.

We  propose the RFM as
a theoretical framework for  linear communication networks.
In this context, the moving particles are data-packets and the chain of sites is a one dimensional set of ordered buffers.
For the RFM with homogeneous link capacities, we
 provide \emph{closed-form}
  expressions for important features including:   steady-state    buffer occupancy,
  network  throughput, and end-to-end delay.
  We use these results to provide
  closed-form expressions for the   hop length and the transmission probability that minimize  the end-to-end delay in ad-hoc multihop linear networks.

The remainder of this paper   is organized as follows.
Section~\ref{sec:pre} reviews the RFM.
Section~\ref{sec:desc} explains how the RFM can be used
to model a linear communication network.
 Section~\ref{sec:main} presents the main results. Section~\ref{sec:app} describes two applications of
  these   results to a multihop  network. Section~\ref{sec:conc} concludes and describes possible directions for future research.
To streamline the presentation, all the proofs are placed in the Appendix.

\section{Ribosome Flow Model}\label{sec:pre}

The RFM is   a set of~$n$ nonlinear first-order ordinary differential equations~(ODEs):
\begin{align}\label{eq:rfm}
                    \dot{x}_1&=\lambda_0 (1-x_1) -\lambda_1 x_1(1-x_2)  \nonumber \\
                    \dot{x}_2&=\lambda_{1} x_{1} (1-x_{2}) -\lambda_{2} x_{2} (1-x_3)  \\&\vdots\nonumber \\
                    \dot{x}_n&=\lambda_{n-1}x_{n-1} (1-x_n) -\lambda_n x_n.\nonumber
\end{align}
This is a \emph{compartmental model}~\cite{comp_surv_1993}, consisting of~$n$ compartments (or sites)
ordered along a linear 1D chain.
Each site  may contain some  material.
(In the case studied here, each site is a buffer
that contains data-packets of fixed length.)
The state-variable~$x_i(t)$, $i\in\{1,\dots,n\}$,
represents  the   occupancy level of site~$i$  at time~$t$.
 The  occupancy   levels are normalized
so that~$x_i(t) \in [0,1]$ for all~$t\geq 0$, with~$x_i(t)=0$ [$x_i(t)=1$]
representing that site~$i$ is completely empty [full] at time~$t$.
Site~$1$  is fed by a  source with \emph{entry rate}~$\lambda_0 > 0$.
Material  flow from site~$i$ to site~$i+1$ depends on a \emph{elongation   rate}~$\lambda_i >0$.
The material  exits at  the end of the chain, that is, from
   buffer~$n$ with \emph{exit  rate}~$\lambda_n > 0$.

To explain the equations of this model, consider the equation
for~$\dot{x}_1$, i.e., the rate of change in occupancy level at site~$1$.
The term~$\lambda_0 (1-x_1) $ is the transition rate into the chain:
as~$x_1$
increases this rate decreases and, in particular, when~$x_1=1$ (i.e., buffer~$1$
is completely full) this rate becomes  zero. The term~$ \lambda_1 x_1(1-x_2)$
is the transition  rate of  material
 from  site~$1$ to the consecutive site~$2$.
This is proportional to the occupancy level at site~$1$,
and also to~$(1-x_2)$. In particular, when~$x_2=1$ this
 becomes zero.  Thus,  as site~$2$ becomes fuller
  the flow from site~$1$ to site~$2$ decreases.
 In this way, the RFM   encapsulates both
the simple exclusion principle and  the  unidirectional movement
  of   TASEP.
The   exit rate from the last site is $R(t):=\lambda_n x_n(t)$.
(In the biological context, this
    represents the
  protein production rate.)

 The state space of the RFM is the~$n$-dimensional unit cube~$C^n:=[0,1]^n$.
 Let~$\Int(C^n)$ denote the interior of~$C^n$, and let~$x(t,a)$ denote the solution of the RFM
 at time~$t$ for the initial condition~$x(0)=a$.
 \begin{proposition}~\cite{margaliot2012stability}\label{prop:stab}
 The RFM admits a unique equilibrium point~$e=e(\LMD) \in \Int(C^n)$,
 and
 for any~$a\in C^n$, $x(t,a) \in \Int(C^n)$ for all~$t>0$, and
 $
            \lim_{t\to \infty}x(t,x_0)=e.
 $
 \end{proposition}
This means that every trajectory   of the
RFM   converges to a steady-state that depends on the rates~$\LMD$.
In particular, $R(t)$  converges to the \emph{steady-state exit  rate}~$R:=\lambda_n e_n$.

The simulation results reported in~\cite{reuveni2011genome} show that
  RFM and    TASEP
  yield similar predictions of steady-state
	  rates. However, recent research results
  on the RFM suggest that expressions for important quantities   are considerably
   simpler than the  corresponding expressions in  TASEP.

Several recent papers analyzed
 the RFM.
In the particular case where  
$
            \lambda_1= \cdots=\lambda_n:=\lambda_c,
$
with~$\lambda_c$ denoting the common value,
the RFM is called  the Homogeneous Ribosome Flow Model~(HRFM).
This model includes only two parameters~$\lambda_0$ and~$\lambda_c$, so its analysis
becomes more tractable~\cite{10.1109/TCBB.2012.120}.
The HRFM with an infinitely-long chain, (i.e. with~$n\to\infty$) was considered in~\cite{infi_HRFM}. There, a simple closed-form
expression for $e_\infty:=\lim_{n\to \infty}e_n$ was derived, as well as explicit
bounds for $|e_\infty-e_n|$ for all~$n\geq 2$.
A closed-loop RFM (representing mRNA circularization)
   has been  studied in~\cite{Margaliot06082013}. It has been shown that the closed-loop system admits a unique globally asymptotically stable equilibrium point.
Ref.~\cite{RFM_entrain}  has shown that
when  the rates~$\lambda_i$ are periodic functions of time, with a common period~$T$, then~\eqref{eq:rfm}
admits a unique periodic solution~$p(t)$, with  period~$T$,
and~$x(t,x_0)\to p(t)$
 for every initial condition~$x_0 \in C^n$. In other words,
 the RFM \emph{entrains}
to periodic excitations in the  rates.
 Ref.~\cite{RFM_concave} has shown that  the steady-state exit  rate
 $R=R(\LMD)$ is a strictly concave function of~$\LMD$ on~$\R^{n+1}_{++}$. This means that the problem of maximizing the steady-state
 exit rate, under an affine constraint on the rates, is a  {convex optimization problem}.

\section{Modeling   Linear Communication Networks using the RFM}\label{sec:desc}
\begin{figure*}[t]
 \centering
 \scalebox{0.75}{\input{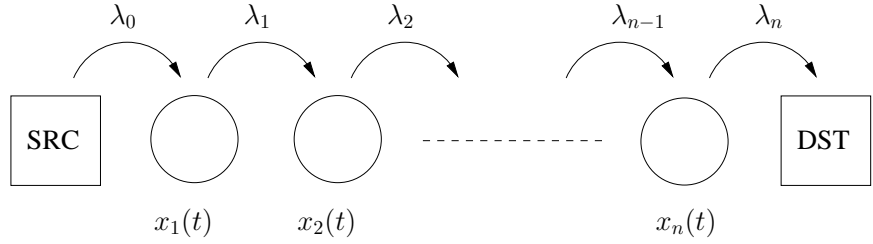}}
\caption{Linear communication network with $n$ relay-nodes modeled using the RFM. SRC [DST]
represents the source [destination] node.
  The circles represent the relays.
 State-variable~$x_i(t)\in[0,1]$ is the normalized   number of data-packets  at buffer~$i$ at time~$t$. The capacity of the link from  site $i$ to site $i+1$ is~$\lambda_i$. The throughput at time~$t$
is~$R(t) =\lambda_n x_n(t)$.}\label{fig:ad_hock_blk}
\end{figure*}

Fig.~\ref{fig:ad_hock_blk} illustrates the general block diagram of a linear communication network with multiple relay nodes modeled using the RFM.
Here the material flowing is data packets, each site corresponds to
 a communication  node that   contains a buffer, and~$x_i(t)$
is the (normalized) occupancy of buffer~$i$ at time~$t$. The parameter~$\lambda_i>0$ is the  capacity
of the link from node~$i$ to node~$i+1$. The network throughput at time~$t$  is~$R(t)$.
The equation~$\dot{x}_i =\lambda_{i-1} x_{i-1} (1-x_{i})
 -\lambda_{i} x_{i} (1-x_{i+1})$ represents a \emph{decentralized}
  flow control algorithm, as the flow through buffer~$i$ depends on the occupancy 
  at buffers~$i-1$ ,$i$, and~$i+1$  only.   

  Prop.~\ref{prop:stab} means that for any  set of strictly positive link capacities
  the flow control algorithm is \emph{stable} in a strong sense, as
the dynamics keeps
every buffer from overflow or underflow at \emph{all} time~$t$ 
(QoS flows are   one example where buffer overflow must be avoided).
Furthermore,  the
    network always converges to a steady state in
   which  the occupancy of buffer~$i$ is~$e_i \in (0,1)$.
    Note  that~$e_i$
    may also be interpreted
 as the probability that site~$i$ is occupied at steady-state.

Recall that   the  \emph{backpressure  algorithm} (see, e.g.,~\cite{TE_stability_queueing,TE_dynamic_server_alloc,BMS_opt_control,GLS_throughput_region,beck_press_neel})
schedules packets  from node~$i$ to node~$i+1$
based on   the differential backlog~$\max\{0,x_i-x_{i+1}\}$.
As~$x_i$ increases and~$x_{i+1}$ decreases,
 the effective flow rate increases. This also happens in the RFM,
 as the flow rate is~$\lambda_i x_i (1-x_{i+1})$.

Note that TASEP is a stochastic, discrete-time, asynchronous model, 
whereas the RFM is a deterministic, continuous-time, synchronous model.
Indeed, in the RFM all the~$x_i$s are updated synchronously.
Conceptually,  the RFM [TASEP] is more suitable for modeling access protocols 
like slotted-ALOHA or CDMA  [TDMA] where the nodes [do not]  transmit at the same time.  

 In the context of communication networks, 
 TASEP corresponds to a 
  chain of buffers, where each buffer can either contain a packet or not. 
  The RFM provides a more flexible framework, as the buffer level~$x_i$ takes values in~$[0,1]$,
  so the buffer can be either empty ($x_i=0$), full ($x_i=1$), half full ($x_i=1/2$), and so on. 
As shown in Section~\ref{sec:main}, analysis of the RFM 
 leads to simpler expressions than those known for TASEP.
 Nevertheless, below 
we also simulate  the 
time-averaged steady-state 
behavior of TASEP
and show that the results agree well with the  analytical formulas
derived for the~RFM.

\section{Analysis of   Linear Communication Networks using the RFM}\label{sec:main}
In this section, we  analyze properties of the RFM that are relevant for   communication networks.
We consider the  special case where all the  link capacities   
  are equal, i.e.~$\lambda_0=\cdots=\lambda_n:=\lambda_c$. We refer to this case as the \emph{totally homogeneous RFM}~(THRFM).


\begin{proposition}\label{prop:occ}
Consider the THRFM with $n$ sites. The steady-state  occupancy level at site~$i$ is
\be\label{eq:ei_simple}
 e_i= \frac{1}{2} + \frac{1}{2} \tan\left(\frac{\pi}{n+3}\right) \cot\left( \frac{(i+1) \pi  }{n+3}\right),\quad i=1,\dots,n.
\ee
 The steady-state throughput is
\be\label{eq:through}
    R=\frac{\lambda_c}{4\cos^2(\frac{\pi}{n+3})}.
\ee
The steady-state delay experienced by a data-packet at node~$i$ is
\be \label{eq:avedel}
 D_i= \frac{2}{\lambda_c}
 \cos^2\left(\frac{\pi}{n+3}\right)
\left (1 +   \tan(\frac{\pi}{n+3}) \cot( \frac{(i+1) \pi  }{n+3})\right) ,
\ee
and the   end-to-end delay is
\be \label{eq:e2edel}
   D_{e2e}:=\sum_{i=1}^n D_i=\frac{2n}{\lambda_c} \cos^2\left(\frac{\pi}{n+3}\right) .
\ee
\end{proposition}

 Eq.~\eqref{eq:ei_simple} implies that the steady-state
buffer occupancies are independent of the homogeneous link capacity~$\lambda_c$ (but~$\lambda_c$ does control the convergence rate
to the steady-state).
Combining~\eqref{eq:ei_simple} with the identity $\cot(x)=-\cot(\pi-x)$  implies the symmetry relation $e_i-\frac{1}{2}=\frac{1}{2} - e_{n+1-i}$, and this gives 
\be\label{eq:sumei}
            \sum_{i=1}^n e_i = n/2.
\ee
This implies that for $n$ odd, $e_{(n+1)/2}=1/2$.
  Eq.~\eqref{eq:ei_simple}
  also  implies that $e_1> \cdots >e_n$. This means that at steady-state, much of the queuing occurs closer to the source node, and preference is given to nodes closer to the destination node. Indeed,
  it was suggested that a scheduling algorithm giving preference to serving links closer to the destination node achieves optimal end-to-end buffer usage in linear networks~\cite{min_buf_usage}.

\begin{figure}[t]
  \begin{center}
  \includegraphics[height=6cm]{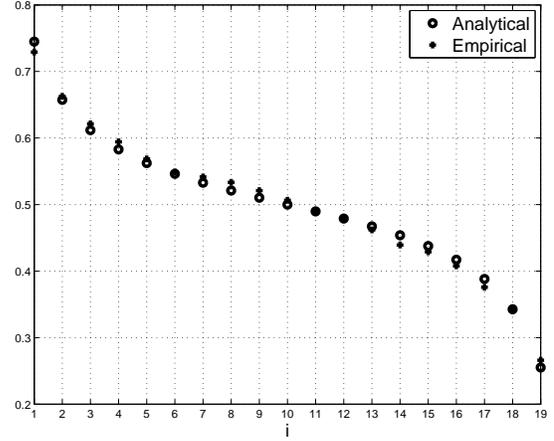}
  \caption{Analytical (O) and empirical (+)   steady-state occupancy level $e_i$  as a function of~$i=1,\dots,19$,
  in a THRFM with $n=19$ nodes.}\label{fig:occu_n19}
  \end{center}
\end{figure}

Fig.~\ref{fig:occu_n19} depicts the~$e_i$s  for a network with $n=19$ nodes.
It may be seen that~$e_i$  decreases with~$i$.
Note also the  symmetry around~$e_{10}=1/2$.
For example,~$ e_{8}-\frac{1}{2}=\frac{1}{2} - e_{12}$.
Fig.~\ref{fig:occu_n19}
also depicts  empirical  occupancy levels  obtained via simulation of
TASEP  with $n=19$ nodes.\footnote{The TASEP simulation used a parallel update mode.
 At each time tick~$t_k$, the sites are scanned from site~$n$ backwards to site~$1$.
 If it is time to hop  and the consecutive site is empty then  the particle advances.
If the consecutive site is occupied the next hopping time,~$t_k+\varepsilon_k$,
is generated. For site~$i$, $\varepsilon_k$
 is  exponentially distributed with parameter~$(1/\lambda_i)$.  The occupancy at each site is averaged through the simulation,
with the first~$350,000$ cycles  discarded in order to obtain  the steady-state value.}

\begin{figure}[t]
  \begin{center}
  \includegraphics[height=6cm]{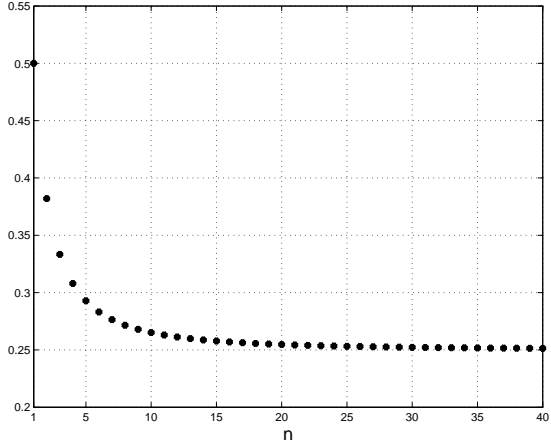}
  \caption{Throughput~$R$ for   link capacity~$\lambda_c=1$ as a function of~$n$. }\label{fig:through}
  \end{center}
\end{figure}

Eq.~\eqref{eq:through} implies that: (1) the steady-state throughput~$R$ is proportional to the homogeneous 
link capacity~$\lambda_c$;
(2)~$R$ is
 upper-bounded by~$\lambda_c/2$ for all~$n$; and
(3) $R$ decreases with the network length~$n$, with~$\lim_{n\to\infty}R=\lambda_c/4$
(see Fig.~\ref{fig:through}).

\begin{figure}[t]
  \begin{center}
  \includegraphics[height=6cm]{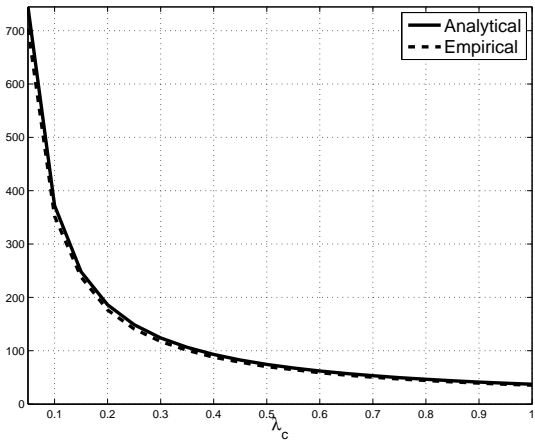}
  \caption{Steady-state end-to-end delay $D_{e2e}$ as a function of the link
	capacity~$\lambda_c$
  in a network with $n=19$ nodes.}\label{fig:delays_n19}
  \end{center}
\end{figure}

Eq.~\eqref{eq:avedel} implies that $D_1> \cdots >D_n$, as the steady-state delay experienced by a data-packet at node~$i$ is proportional to the occupancy level at that node. The end-to-end delay~$D_{e2e}$ is inversely proportional to the link capacity~$\lambda_c$,
and   increases more or less linearly with the network length~$n$. In particular,~$D_{e2e}\approx
 {2 n} / {\lambda_c}$ for large values of~$n$.
Fig.~\ref{fig:delays_n19} depicts $D_{e2e}$ as a function of $\lambda_c$ for a THRFM with $n=19$. It also shows the empirical values of $D_{e2e}$ obtained via simulating TASEP and
  averaging the end-to-end delays   of data-packets
  (after discarding the first~$350,000$ cycles).

\section{Application to Multihop Networks}\label{sec:app}

Ad-hoc multihop networks and their military and civilian applications
are attracting considerable interest, yet
  there remain significant theoretical
   challenges in the performance analysis of these networks~\cite{beyond_shannon,multihop_theory,rethinking_info},
   and thus in their systematic design. The decentralized,
    multiagent network structure
     and the possibility for
    multihop routing  make
       the analysis of fundamental communication metrics, including throughput, latency and capacity  quite challenging.
In 
multihop networks,  data-packets may be routed over many short hops or over a smaller number of long hops.
An important research topic    is determining various parameters, e.g. the
 hop length, that optimize some performance metric, e.g. minimize the end-to-end delay.
(Of course, in general there could be additional   factors   impacting the decision between short and long hops~\cite{ad_hoc_long_hops}.)
In this section we show how to use the analysis results on the~THRFM described above
 to systematically  address this type of optimization problems.

We assume that all the nodes in the network use the same
channel. 
The nodes are  equally spaced  along the communication system,
 with a separation~$\ell$  between any two adjacent nodes.
Attenuation in the channel is
modeled as the product of a Rayleigh fading component and a
  large-scale path loss component with exponent~$\gamma$.
  The channel noise is taken to be additive white Gaussian noise (AWGN) with power spectral density
$N_0$. We define the transmission from node $i$ to  node $i+1$
to be successful if the (instantaneous) signal-to-interference-and-noise ratio~(SINR) at   node~$i+1$ is greater than a predetermined
threshold~$\Theta$. The probability of a successful transmission is thus~$p_s:=\text{Pr}( \text{SINR} >\Theta)$.

\subsubsection{Optimal Hop Length at SNR Limit Regime}

Consider an ad-hoc multihop linear  network where the main effect on a successful transmission is the channel attenuation and noise. This happens, for example, when the nodes are relatively far  so that the effect of the spatial reuse is negligible relative to the channel noise. In this regime,
\be  \label{eq:SNR}
      p_s=\text{Pr}( \text{SNR} >\Theta) = \exp(-\Theta N_0 l^\gamma).
\ee

We   model this using the THRFM with link capacities~$\lambda_c =r p_s$, where~$r>0$.
For simplicity, we take~$r=1$.
The end-to-end delay in the THRFM given in~\eqref{eq:e2edel}
then becomes
\be \label{eq:e2edelwithlamc}
   D_{e2e}= 2  n  \exp(\Theta N_0 l^\gamma) \cos^2\left(\frac{\pi}{n+3}\right) .
\ee
The closed-form expression for~$D_{e2e}$ allows us to address various optimization problems.
For example, consider the problem of finding the 
 hop length that minimizes the end-to-end delay.
 Let $ml$ denote a hop of $m$ steps of unit length~$l$. Replacing $l$ with $ml$ and $n$ with $n/m$ in~\eqref{eq:e2edelwithlamc} yields
 \be \label{eq:e2edfii}
    D_{e2e}= \frac{2n}{m}  \exp(\Theta N_0 (ml)^\gamma) \cos^{2}\left(\frac{\pi   }{(n/m)+3}\right).
\ee
In particular, for~$n/m \gg 1$,
 \be\label{eq:e2esim}
   D_{e2e}\approx \bar D_{e2e} : = \frac{2n}{m}   \exp(\Theta N_0 (ml)^\gamma)  .
\ee
This expression shows that~$m$ affects~$D_{e2e}$ in two different ways.
On the one-hand, increasing~$m$ decreases the total length of the chain and thus
decreases the delay. On the other-hand, increasing~$m$ increases the separation between the communicating nodes. 
This decreases the probability of a successful transmission thus increasing the delay. 

\begin{proposition} \label{mopt}
The  hop length that minimizes the end-to-end delay~$\bar D_{e2e} $  in the  SNR limit regime is
either~$  \lceil m^*  \rceil$ or~$\lfloor m^* \rfloor$, where
\be\label{eq:mstar}
    m^*: =  \frac{1}{l} \left(\frac{1}{\gamma\Theta N_0}\right)^{ {1}/\gamma},
\ee
and for this value, the optimal end-to-end delay is
\[
\bar D_{e2e}^* = {2}  n l    \left( \gamma\Theta N_0 \right)^{ {1}/\gamma}  \exp(1 /\gamma).
\]
\end{proposition}

\begin{figure}[t]
  \begin{center}
  \includegraphics[height=6cm]{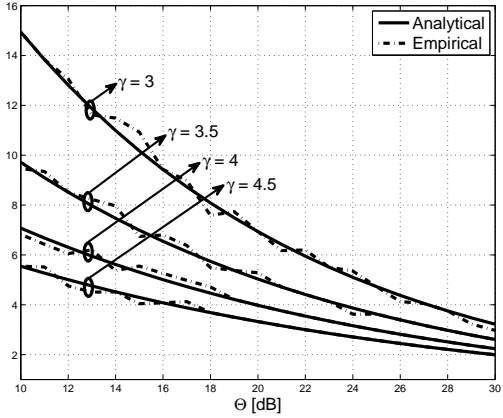}
  \caption{  Optimal hop length~$m^*$  as a function of the SNR threshold~$\Theta$ and the path loss exponent~$\gamma$ for a network with $n=200$ nodes.}\label{fig:opt_m}
  \end{center}
\end{figure}

Eq.~\eqref{eq:mstar} implies that $m^*$ is monotonically decreasing with $\Theta$ and~$N_0$. This is reasonable,
 as an increase in any of these parameters decreases the effective SNR at the receiving node, which in turn decreases the probability of a successful transmission. This can be compensated by decreasing the hop length.
Fig.~\ref{fig:opt_m} depicts the analytical and empirical values of~$m^*$ as a function of~$\Theta$ and~$\gamma$ for~$\ell=1$ and~$N_0=-50$ dBW/Hz.
 The empirical values were obtained by simulating several TASEPs, 
 each one with~$200/m$ sites and with an
 exponentially distributed next time to hop with parameter~$\lambda_c =p_s$ as given in~\eqref{eq:SNR},
 but with~$\ell$ replaced by~$m\ell$.

\subsubsection{Optimal Transmission Probability at SIR Limit Regime}
Consider now an ad-hoc multihop wireless network
where 
   the main effect on a successful transmission is the transmission interference.
Let~$q$ 
denote the contention probability, i.e. the probability that a node transmits at each time slot.
 In this case, 
 \be  \label{eq:SIR}
    p_s  \approx \exp({-qc}/{2}),
\ee
where $c:=\frac{\pi\Theta^{\frac{1}{\gamma}}}{\sqrt{ \gamma / 2 }}-1$ (see~\cite{DBLP:journals/twc/Haenggi09}).

A natural question is how  to determine the contention probability that \emph{minimizes}~$D_{e2e}$.
To address this, we model this scenario as a
  THRFM with~$\lambda_c = q p_s$. 
Substituting this in~\eqref{eq:e2edel} yields
\be\label{eq:d_new}
   D_{e2e}=(v /q) \exp(qc/2),
\ee
where~$v:=2 n  \cos^2\left(\frac{\pi}{n+3}\right)$.
This expression shows that~$q$ affects~$D_{e2e}$ in two different ways.
On the one-hand, increasing~$q$ increases the link capacity
 and thus decreases the delay. On the other-hand, increasing~$q$ increases the interference.
This decreases the probability of a successful transmission thus increasing the delay.


\begin{proposition} \label{prop:lmabdaopt}
The   contention probability that minimizes
 the end-to-end delay in the SIR limit regime is
\be  \label{eq:qopt}
    q^*= \min\{1,2/c\}.
\ee
\end{proposition}

\begin{figure}[t]
  \begin{center}
  \includegraphics[height=6cm]{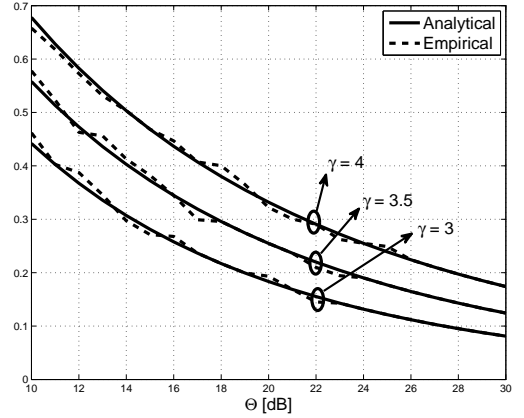}
  \caption{Optimal transmission probability $q^*$  as a function of the   threshold~$\Theta$ and the path loss exponent~$\gamma$.}\label{fig:opt_q}
  \end{center}
\end{figure}

Fig.~\ref{fig:opt_q} depicts the analytical and empirical optimal contention probability $q^*$ as a function of $\Theta$ and $\gamma$.
Note that~$q^*$ is inversely proportional to both~$\Theta$ and to~$\gamma$.

\section{Discussion}\label{sec:conc}
TASEP is an important model in non-equilibrium statistical mechanics that has been used to study numerous biological and artificial multiagent systems
 including multihop wireless networks~\cite{DBLP:journals/tmc/SrinivasaH12,liutowards,conf/icccn/SrinivasaH10}.
The   RFM is a ``deterministic substitute'' of   TASEP that is highly
 amenable to mathematical analysis.
In this paper, we proposed the RFM as a new theoretical framework for modeling and
analyzing linear communication networks. This has several advantages.
The dynamics of the RFM 
guarantees that   the buffer levels do not overflow or underflow for \emph{all} time~$t$,
and converge to a unique steady-state level. 
 For the particular case of the THRFM, another important
  advantage   is that  simple 
 closed-form expressions for important network metrics such as the  steady-state 
throughput and end-to-end-delay exist.
These expressions can be used to select optimal parameters, for example, the optimal hop length
in a multihop network. 

For the more general case of an RFM with non-homogeneous rates closed-form expressions for the
network metrics are not known. However, they can be easily computed   in an efficient and numerically stable way. Indeed, it has been shown in~\cite{RFM_concave} that  the~$(n+2)\times(n+2)$ matrix
\begin{align}\label{eq:rfm_matrix_A}
 \small
                A := \begin{bmatrix}
 0 &  \lambda_0^{-1/2}   & 0 &0 & \dots &0&0 \\
\lambda_0^{-1/2} & 0  & \lambda_1^{-1/2}   & 0  & \dots &0&0 \\
 0& \lambda_1^{-1/2} & 0 &  \lambda_2^{-1/2}    & \dots &0&0 \\
 & &&\vdots \\
 0& 0 & 0 & \dots &\lambda_{n-1}^{-1/2}  & 0& \lambda_{n }^{-1/2}     \\
 0& 0 & 0 & \dots &0 & \lambda_{n }^{-1/2}  & 0
 \end{bmatrix}
\end{align}
has real and distinct eigenvalues:
$
            \zeta_1< \zeta_2< \dots < \zeta_{n+2},
$
and that the steady-state throughput in the~RFM is~$	R=  \left(\zeta_{n+2}\right ) ^{-2}$.
This means that~$R$, and then every other steady-state metric, can be determined  using numerical
algorithms for calculating the eigenvalues of symmetric (tridiagonal) matrices.

The  link capacities  in a communication network
must often be constrained, e.g. to reduce interference or because
of energy constraints~\cite{Throughput_max_2014}.
A natural question is how to maximize the throughput
subject to  a constraint on the 
capacities. 
In the linear communication
network modeled using the RFM, this leads
 to the following optimization problem:
\begin{Problem}\label{prob:max}

Fix~$b, w_0,w_1,\dots,w_n  >0$.  Maximize~$R =R(\LMD)$, with respect to its parameters $\LMD$, subject to the constraints:
\begin{align}\label{eq:constraint}
\sum_{i=0}^n w_i\lambda_i  & \leq  b, \\
\LMD &\geq 0.\nonumber
\end{align}
\end{Problem}
The values~$w_i$ can be used to provide different weighting to the different link capacities.
 It has been shown in~\cite{RFM_concave}
 that~$R$ is a strictly concave function of the   $\lambda_i$s, and since
 the constraint on the rates is affine, Problem~\ref{prob:max} is a \emph{convex optimization problem}. Thus, 
   the solution $\lambda_i^*$, $i=0,\dots,n$, is unique and can be found using efficient numerical algorithms that scale well with $n$.
 
\begin{figure}[t]
  \begin{center}
  \includegraphics[height=6cm]{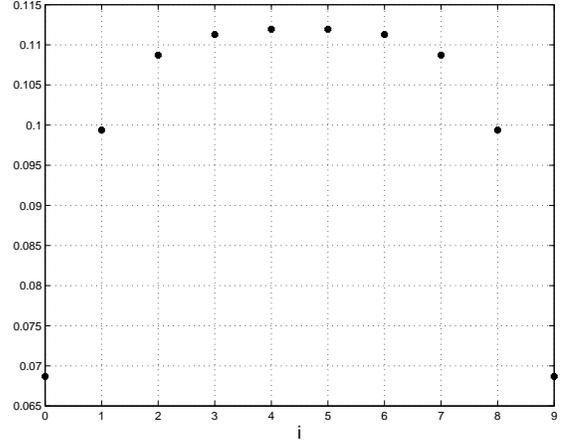}
\caption{Optimal link capacities~$\lambda_i^*$ as a function of $i$ for a linear
communication network with $n=9$ nodes.}\label{fig:rfm_n9_l}
  \end{center}
\end{figure}

Fig.~\ref{fig:rfm_n9_l} depicts the optimal link capacities~$\lambda_i^*$, $i=0,\dots,9$, for a communication network with $9$ relay nodes and the  constraint~\eqref{eq:constraint} with~$w_0=\dots=w_n=b=1$.  It may be seen
  that the~$\lambda_i^*$s are symmetric, i.e.
  $\lambda_i^*=\lambda_{9-i}^*$, and that
   they increase towards the center of the network. In other words,
   the capacities near the center of the network have a higher effect on the throughput than
   those near the ends of the network.

Another interesting research topic is to model and analyze
   more complex network topologies, comprising of multiple flows that share common nodes, using a set of \emph{interconnected} RFMs.

\section*{Acknowledgments}
We thank Sunil Srinivasa
for helpful discussions and for providing us   Matlab programs simulating the system in~\cite{DBLP:journals/tmc/SrinivasaH12}. We thank Boaz Patt-Shamir for helpful discussions.

\section*{Appendix: Proofs}
{\sl Proof of Prop.~\ref{prop:occ}.}
It has been shown in~\cite{10.1109/TCBB.2012.120}
that for an HRFM with $m$ sites and~$\lambda_0 \to \infty$ the
equilibrium  point~$e=\begin{bmatrix} e_1, \dots, e_m\end{bmatrix}'$ satisfies
$
 e_i=\frac{d_i}{2\cos(\frac{\pi}{m+2})}$,
 $i=1,\dots,m$,
where~$d_i:=\cos (\frac{\pi}{m+2} )+\sin (\frac{\pi}{m+2} )\cot ( \frac{i \pi  }{m+2} )$.
In particular,~$e_1=1$. Since~$\lambda_1=\dots=\lambda_n=\lambda_c$,~\eqref{eq:rfm}
implies that the steady-state satisfies
$\begin{bmatrix} e_1 & e_2 &\cdots &e_m\end{bmatrix}=\begin{bmatrix} 1 & \tilde e_1 &\cdots &\tilde e_{m-1}\end{bmatrix}$, where~$\begin{bmatrix}  \tilde e_1 &\cdots &\tilde e_{m-1}\end{bmatrix}$ is the equilibrium
of a THRFM with dimension~$m-1$.
Taking~$m=n+1$ yields~\eqref{eq:ei_simple}.
At steady-state, the exit rate is
$R=\lambda_n e_n=\lambda_c e_n$, and this is equal to the rate of packet flow
across each node. Combining this with~\eqref{eq:ei_simple} yields~\eqref{eq:through}.
By  Little's Theorem~\cite{Kleinrock:1975:TVQ:1096491},
  the delay at node~$i$ at steady-state is~$D_i=e_i/R$. Combining this
 with~\eqref{eq:through} and~\eqref{eq:ei_simple} yields~\eqref{eq:avedel}.
The end-to-end   delay is
$    \sum_{i=1}^{n}D_i=(1/R)\sum_{i=1}^{n}e_i$,
and using~\eqref{eq:sumei}  and~\eqref{eq:through} yields~\eqref{eq:e2edel}.~$\IEEEQED$

{\sl Proof of Prop.~\ref{mopt}}.
Differentiating~$\bar D_{e2e}$
  with respect to~$m$ yields
$
 \frac{\partial}{\partial m} \bar D_{e2e}=
\frac{2 n}{m^2} \exp(N_0 \Theta  (l m)^{\gamma }) \left(\gamma  N_0
   \Theta  (l m)^{\gamma }-1\right)  ,
 $ and
$
\frac{\partial^2}{\partial m^2} \bar D_{e2e}=\frac{2n}{m^3}\exp(z)(\gamma^2z^2+z\gamma(\gamma-3)+2),
$
where $z:=\Theta N_0 (ml)^\gamma$.
It is straightforward
 to verify that $\frac{\partial^2}{\partial m^2}\bar D_{e2e}>0$ for all $m>0$ and $\gamma \ge1$, thus for all practical cases  $\bar D_{e2e}$   is a strictly convex function of $m$,
 and the unique global minimum $m^*$ can be found by equating $ \frac{\partial}{\partial m} \bar D_{e2e}$ to zero, which yields~\eqref{eq:mstar}  (of course, the relevant  value is either~$\lceil m^* \rceil$
 or~$\lfloor m^* \rfloor$, as the hop length must be an integer).
Substituting~$m^*$ in~\eqref{eq:e2esim} completes the proof.~$\IEEEQED$

{\sl Proof of Prop.~\ref{prop:lmabdaopt}.}
Differentiating~\eqref{eq:d_new}  yields
 $
\frac{\partial}{\partial q} D_{e2e} =v\exp(cq/2)\frac{ c q-2 }{2q^2} ,
$
and
$
\frac{\partial^2}{\partial q^2} D_{e2e}= v \exp(cq/2) \frac{ f(cq)}{4q^3},
$
where $f(z):=z^2-4z+8$. Since~$v>0$ and~$f(z)>0$ for all $z$, it follows that $\frac{\partial^2}{\partial q^2} D_{e2e}>0$ for all $q>0$, so~$D_{e2e}$ is a strictly convex function of $q$, and the unique global minimum $q^*$ can be found by equating~$\frac{\partial}{\partial q} D_{e2e}$ to zero. This yields~$q^*=2/c$. If~$c\geq 2$ then~$q^*=2/c \leq 1$, so this is a valid minimizer (recall
that~$q\in(0,1]$). If~$c<2$ then~$2/c >1$, so the valid minimum is~$q^*=1$.~$\IEEEQED$

  \newpage

\bibliographystyle{IEEEtran}  
\bibliography{wlanrfm}
\end{document}